\documentclass[epj,referee]{svjour}
\usepackage{epsfig,amssymb,amsmath}
\usepackage{color}
\begin{document}
\title{Diffusing opinions in bounded confidence processes}
\author{M. Pineda\inst{1}, R. Toral\inst{2} \and E. Hern\'andez-Garc\'{\i}a\inst{2}
}
\institute{Center for Nonlinear Phenomena and Complex Systems,
Universit\'e Libre de Bruxelles, Code Postal 231, Campus Plaine,
B-1050 Brussels, Belgium. \and IFISC, Instituto de F\'{\i}sica
Interdisciplinar y Sistemas Complejos (CSIC-UIB), Campus UIB, E-07122
Palma de Mallorca, Spain.}
\date{Received: date / Revised version: date}

\abstract{ We study the effects of diffusing opinions on the
Deffuant {\sl et al.} model for continuous opinion dynamics.
Individuals are given the opportunity to change their opinion,
with a given probability, to a randomly selected opinion inside
an interval centered around the present opinion. We show that
diffusion induces an order-disorder transition. In the
disordered state the opinion distribution tends to be uniform,
while for the ordered state a set of well defined opinion
clusters are formed, although with some opinion spread inside
them. If the diffusion jumps are not large, clusters coalesce,
so that weak diffusion favors opinion consensus. A master
equation for the process described above is presented. We find
that the master equation and the Monte-Carlo simulations do not
always agree due to finite-size induced fluctuations. Using a
linear stability analysis we can derive approximate conditions
for the transition between opinion clusters and the disordered
state. The linear stability analysis is compared with Monte
Carlo simulations. Novel interesting phenomena are analyzed. }
\maketitle

\section{Introduction}
In a community, opinions evolve due to affinities and conflicts
between mutually interacting individuals. These interactions
lead to collective states, where either a majority of
individuals adopt a similar opinion (consensus) or a number of
opinion groups (clusters) arise. In many situations, the
dynamics of this complex collective behavior goes beyond
specific individual attributes and seems to be well
characterized by quantities like statistical distributions and
averages. This explains why, in the last years, the
understanding of opinion formation using tools and techniques
borrowed from nonlinear and statistical physics has become a
topic of interest for physicists. In particular, several models
have been developed to reproduce the basic elements that drive
the processes of opinion evolution \cite{castellano}. These
models can be classified in two broad groups: on the one hand,
discrete, Ising-type, models where opinions can only adopt a
finite set of integer values \cite{galam,sznajd}; on the other,
continuous opinion models where opinions can vary continuously
in a finite interval
\cite{schweitzer,lorenz,deffuant1,deffuant2,deffuant3,krause1,markowich}.

In the context of continuous opinion dynamics a model
introduced by Deffuant and collaborators has received much
attention recently \cite{deffuant1}. In this model individuals
meet in random pairwise encounters and then converge to a
common opinion only if their respective opinions prior to the
encounter differ less than some given amount, in a kind of {\sl
bounded confidence} mechanism \cite{axelrod,granovetter}. After
some transient evolution, this leads to final states in which
either full consensus is reached or the population splits in a
finite number of clusters such that all individuals in one
cluster share the same opinion. However, we believe that such
dynamics misses the fact that the behavior of an individual in
a society does not depend only on the influence of people with
similar opinions. There are many additional factors, both
personal and external, which also cause opinion changes. These
factors may introduce some degree of discrepancy within
otherwise well defined opinion clusters and should be modeled
somehow. In a recent paper \cite{pineda} we analyzed the
effects induced by a modification of the model in which we
allow individuals to change their opinion, with a given
probability, to a randomly chosen value in the whole opinion
space. One possible interpretation of this modification (which,
from the technical point of view, can be understood as some
sort of noise acting upon the dynamics) is that individuals
keep at all times a basal opinion to which they return from
time to time, no matter what the environment tells them to do,
modelling in a crude way the intrinsic free-will of human
decisions. We have shown \cite{pineda} that this modification
of the dynamical rules can induce new and interesting
phenomena, such as noise-induced bistability, and that the
model displays interesting and non-trivial finite-size effects
that need to be considered in some detail.

In this paper, we continue our analysis of the uncertainty
factors present in human decisions by considering a
modification of the model in which the opinion of an individual
can jump to a new value chosen randomly inside an interval of
finite width centered around the current opinion. In other
words, in addition to the bounded confidence mechanism, the
opinions execute a random-walk motion or diffusion in opinion
space. When the typical size of the jumps is sufficiently large
we expect this model to behave as the previous free-will model
discussed in \cite{pineda}, but when it is small a distinct
behavior will be obtained.

We analyze which aspects of the Deffuant {\sl et al.} model are
robust against the introduction of this diffusion process. The
analysis, based upon Monte Carlo simulations as well as on
numerical integrations of the corresponding master equation,
reveals indeed a dependence on the size of the random jumps.
For small jumps, we have found that the center of mass of each
opinion cluster executes a random walk with an effective
diffusion coefficient equal to the one corresponding to a
single individual divided by the number of individuals in the
cluster. The interplay between interactions and random walk
induces a coarsening process that leads to the formation of a
single large cluster at very long times. When large jumps are
allowed, the clusters tend to stay in relatively well defined
positions. We have observed, in a way similar to the model in
\cite{pineda}, regions of bistability in which fluctuations can
take the system from one state to another and back. We will
show that Monte Carlo simulations do not necessarily agree with
the results of the master equation because of the inherent
finite-size-induced fluctuations. Finally, our studies reveal
that there is a region in parameter space in which the system
becomes disorganized and cluster formation does not occur. We
provide a linear stability analysis that qualitatively
reproduces this order-disorder transition. Analytical results
are compared with Monte Carlo simulations using the so-called
cluster coefficient which aims to characterize the existence of
clusters.

This paper is organized as follows: the model and main results
coming from Monte Carlo simulations and numerical integration
of the master equation are presented in section
\ref{sec:noise}; in section \ref{sec:orderdisorder} we use the
so-called group coefficient to characterize the order-disorder
transition that appears in the model and we apply a linear
stability analysis to derive approximately the critical
parameter values for the formation of opinion clusters; summary
and conclusions are presented in section \ref{sec:conclusions}.

\section{A continuous-opinion dynamics model with bounded opinion jumps}
\label{sec:noise} The original version of the Deffuant {\sl et
al.} model \cite{deffuant1} considers a population with $N$
individuals. The opinion $x_n^i$ on a given topic that
individual $i$ has at time-step $n$ is a real variable in the
interval $[0,1]$. One also assumes that the initial values
$x_0^i$ for $i=1,\dots,N$ are randomly distributed in this
interval. A bounded confidence mechanism is introduced to
reflect that individuals interact, discuss, and modify their
opinions: at time-step $n$ two individuals, say $i$ and $j$,
are randomly chosen; if their opinions satisfy
$|x_n^i-x_n^j|<\epsilon$, they converge to the common value:
\begin{equation}
\label{eq:rule}
x_{n+1}^{i}=x_{n+1}^{j}=\frac{x_{n}^{j}+x_{n}^{i}}{2},
\end{equation}
otherwise they remain unchanged. Whether the opinions have been
modified or not, time increases $n\to n+1$. It is customary to
introduce the time variable $t=n\Delta t$, where $\Delta
t=1/N$, measuring the number of opinion updates per individual,
or number of Monte-Carlo steps (MCS). As a consequence of the
iteration of this dynamical rule, the system reaches a static
final configuration, which depends on the {\sl confidence
parameter} $\epsilon$ taking values between 0 and 1. Starting
from uniformly distributed random values for the initial
opinions, the typical realization is that for $\epsilon \geq
0.5$ the system evolves to a state of consensus where all
individuals share the same opinion and that, decreasing
$\epsilon$, the population splits into opinion clusters
separated by distances larger than $\epsilon$
\cite{lorenz,redner}.

The new ingredient we add to the dynamics is that individuals
can perform random jumps in their opinions. More specifically
the dynamical rules are modified as follows: at time step $n$
the dynamical rule Eq. (\ref{eq:rule}) applies with probability
$1-m$; otherwise a randomly chosen individual $i$ changes the
opinion to a new value $x_{n+1}^i$ randomly chosen from the
interval $(x^{i}_{n}-\gamma,x_n^i+\gamma)$. The parameters
$\gamma\in[0,1]$ and $m\in[0,1]$ determine, respectively, the
width and frequency of random jumps. When $\gamma$ is large
enough, we expect that the behavior of this model will approach
the one of our previous free-will model \cite{pineda}, in which
noisy jumps occurred to random locations in opinion space.
Since the variance of the opinion jumps is
$\sigma^2=\gamma^2/3$, each individual would behave as a random
walker with diffusion coefficient $D=m\gamma^2/3$ if
interactions and boundaries were absent. Note, however, that it
is possible from the model rules that opinions leave the
bounded opinion space $[0,1]$. In order to assure that opinions
stay inside this interval, we need to implement proper boundary
conditions to the dynamics. A simple and mathematically
convenient choice is to use {\sl periodic boundary conditions},
so that the opinion interval is considered to be wrapped on a
circle. In this case, opinions that go away a certain distance
to the left of the extreme opinion $0$ or to the right of the
extreme opinion $1$ are injected throughout the opposite
extreme by a similar amount. In real situations it is unlikely
for individuals with extreme opinions to change their opinions
so drastically. Thus, we will use more realistic, {\sl
adsorbing} boundary conditions, in which opinions that try to
go away towards the left or towards the right of the interval
$[0,1]$ are set to $0$ or $1$, respectively. Nevertheless, for
mathematical simplicity periodic boundary conditions will be
considered in some particular cases as properly mentioned.

\subsection{Monte Carlo simulations}
\label{sec:MCS} 
\begin{figure}
\centering
\mbox{\epsfig{file=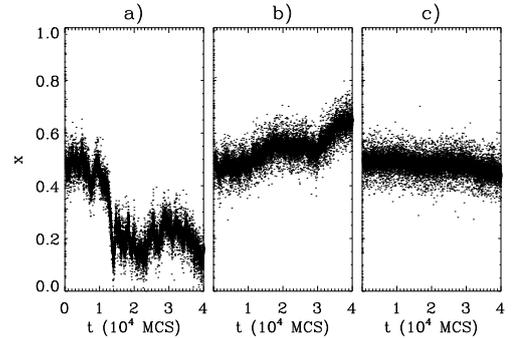,width=0.9\linewidth,clip=,angle=0}}
\hfill
\caption{\small \small Time series of the opinion distributions
at three values of
$N$. $m=0.1$, $\gamma=0.1$, and $\epsilon=0.28$.
(a) At $N=100$ the cluster moves around the whole opinion space.
(b) The cluster at $N=1000$ moves less. (c) At $N=10000$ the cluster remains close to $0.5$.
Simulations are performed with homogeneous initial conditions and adsorbing boundary conditions.
Although the simulations were done with 100, 1000 and 10000 agents, here we plot in all panels
only 100 of them, and at intervals of 100 MCS, to avoid saturation of the plots.}
\label{fig1}
\end{figure}

We present in this subsection the main results
obtained from Monte Carlo simulations for a system with a
finite population of $N$ individuals and adsorbing boundary
conditions. As in previous studies, we assume that the initial
condition represents a uniform distribution in opinion space
interval $[0,1]$. In the original Deffuant {\sl et al.}  model
($m=\gamma=0$), Monte Carlo simulations show that for
$\epsilon>0$ and starting with homogeneous initial conditions
the system either reaches a final state of perfect consensus or
splits into a finite number of clusters such that all
individuals in one cluster have exactly the same opinion
\cite{deffuant1}. This picture changes when diffusion is
introduced in the way described above. First, for $m>0$ and
$\gamma>0$, clusters still are formed but individuals in a
single cluster do not have exactly the same opinion but there
is some dispersion among them. Second, when $\gamma$ is small
the center of mass of each cluster performs a random walk
through the  opinion space. As a consequence, opinion clusters
start to collide and merge in a coarsening process that leads
finally to a single surviving large group at very long times.
To visualize these interesting behaviors, Fig.~\ref{fig1} shows
time series of the opinion space for a value of $\epsilon$ such
that only one big cluster is formed. Dispersion around the
center of mass remains approximately independent of $N$, but it
is seen that the center of mass $X_{cm}$ of the cluster behaves
as a random walker in opinion space, with smaller displacement
for larger $N$. We can understand this motion by noting that
the  binary opinion interactions do not
change the position of the center of mass, which then moves only because of the
diffusive motions of the opinions. We can thus write for the
center of mass $X_{cm}$ at time $t+1$ (i.e., after the number
of steps has been increased from $n$ to $n+N$) as its position
at time $t$ plus the motion induced by the individual jumps
occurring in that time interval (let us say that there have
been $N_1$ of them):
\begin{equation}
 X_{cm}(t+1) \equiv \frac{1}{N}\sum^{N}_{i=1}x^{i}(t+1) =
X_{cm}(t) + \frac{1}{N}\sum^{N_1}_{i=1} \xi^{i}(t).
\end{equation}
$\xi^{i}(t)$ are independent random jumps of zero mean and variance
$\sigma^2=\gamma^2/3$. Reordering, and taking squares and mean
value one obtains
\begin{eqnarray}
\left<[X_{cm}(t+1)-X_{cm}(t)]^{2}\right> & = &
\left<\left(\frac{1}{N}\sum^{N_1}_{i=1} \xi^{i}(t)\right)^2\right> \nonumber  \\
& = & \frac{m N}{N^2}\sigma^2=\frac{m}{N}\frac{\gamma^2}{3},
\label{CM}
\end{eqnarray}
where we have used that the expected value of $N_1$ is $mN$.
This implies that the cluster's center of mass experiences a
random walk with an effective diffusion coefficient $D_{cm}$
equal to the single-individual's opinion one divided by the
number of individuals in the cluster $D_{cm}=D/N$. Fig.~\ref{fig1} illustrates cluster random walks for three different population sizes. For isolated clusters (i.e. separated by distances larger than $\epsilon$) the arguments
above are only valid when boundary effects are unimportant,
which means that clusters remain sufficiently far from the
interval extremes (at distances larger than $\gamma$) so that
jumping particles can not reach the boundaries. For larger
$\gamma$ or $\epsilon$ particles will feel the boundaries more
easily and we expect cluster mobility to be much reduced by
boundary effects. 

\begin{figure}
\centering
\mbox{\epsfig{file=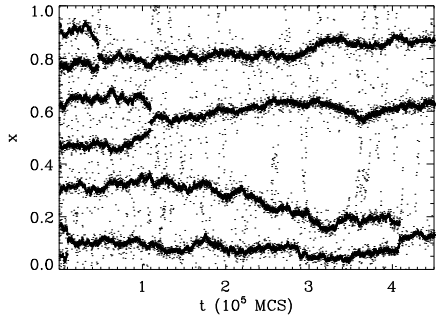,width=0.9\linewidth,clip=,angle=0}}
\hfill
\caption{\small Time series of the opinions with
$\epsilon=0.05$, $m=0.01$, $\gamma=0.04$, and $N=1000$. To avoid saturation of the plot, only
100 agents are shown, and at intervals of 450 MCS. The clusters perform random walks and successive
merging of clusters occurs after collisions. Finally, at very long times, a state containing a single cluster will dominate.
Simulations are performed with homogeneous initial conditions and adsorbing boundary
conditions.}\label{fig2}
\end{figure}

From Monte Carlo simulations, it is also seen that cluster
formation always occurs for small values of $\epsilon$,
$\gamma$, and $m$. But, in contrast with the diffusionless
model, coarsening is observed for small $\gamma$.
Figure~\ref{fig2} shows successive merging of clusters,
occurring after the collisions which arise because of the
diffusive wandering of clusters (by {\sl collision} we mean
that clusters become closer than $\epsilon$, so that they
interact). We expect that in this regime, at very long times, a
state containing a single cluster would be the final regime.

\begin{figure}
\centering
\mbox{\epsfig{file=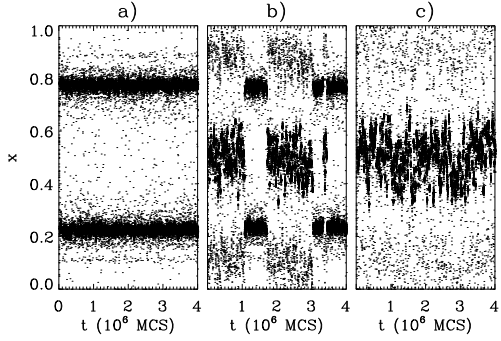,width=0.9\linewidth,clip=,angle=0}}
\hfill
\caption{\small Time series of the opinion distributions at three
values of $\epsilon$ and $m=0.01$, $\gamma=0.4$, and $N=1000$.
To avoid saturation of the plot, only
100 agents are shown, and at intervals of 1000 MCS. (a)
At $\epsilon=0.28$ the system appears polarized in two opinion
clusters. (c) A single major cluster with two lateral minor
clusters is observed at $\epsilon=0.35$. (b) At $\epsilon=0.316$
the systems fluctuates between these two states.
Simulations are performed with homogeneous initial conditions
and adsorbing boundary conditions.}\label{fig3}
\end{figure}

The behavior is different at large values of
$\gamma$: in addition to the expected reduced wandering of the
clusters, which tend to stay at relatively well defined
positions, we see that there is no tendency to reducing the
number of clusters (See Figs. \ref{fig3}a and c), so that a
pattern of opinions is established. The pattern of clusters is
approximately periodic when $\epsilon \ll 1$. This is similar
to the behavior of the model in \cite{pineda}. In fact, for
$\gamma$ of the order of system size the jumps are of global
size so that there should be no difference between both
models. As in \cite{pineda}, a remarkable fact occurs in which
one can find regions of bistability: inside these regions, the
inherent fluctuations arising from the finite number of
particles take the system from one state to another and back.
The sort of transitions between steady states is observed for
instance, in Monte Carlo simulations at $\epsilon=0.316$.
Figure~\ref{fig3}b shows multiple jumps between a state of two
big opinion cluster and another state of a big cluster with two
smaller ones near the edges of the opinion interval.

\subsection{Master equation approach}
In this subsection we analyze the master equation description
of the process introduced above. Following standard arguments
(see for example \cite{pineda}) one finds the master equation
for the probability density function $P(x,t)$ of an individual
opinion $x$ at time $t$
\begin{eqnarray}
  \frac{\partial P(x,t)}{\partial t} & = & (1-m)[4\int_{|x-x_{2}|<\epsilon/2}dx_{2}P(2x-x_{2},t)P(x_{2},t) \nonumber \\
  & & - 2P(x,t)\int_{|x-x_{2}|<\epsilon}dx_{2}P(x_{2},t)] \nonumber  \\
  & & + m\left[G(x,t)-P(x,t)\right].
 \label{eq:defuantdiff}
\end{eqnarray}

The term proportional to $(1-m)$ is the one coming from the
original rules of the Deffuant {\sl et al.} model
\cite{lorenz,redner}, whereas the one proportional to $m$
describes the random jumps. The function $G(x,t)$ is, for the
case of adsorbing boundary conditions:
\begin{equation}
G(x,t)=
\left\{ \begin{array}{ll}
 \delta(x) \int_0^{\gamma} dx^{'}\frac{\gamma-x^{'}}{2\gamma}P(x^{'},t)\\
+\int_0^{x+\gamma} \frac{dx^{'}}{2\gamma}P(x^{'},t), & \mbox{if } x\leq\gamma, \\
\\
\int_{x-\gamma}^{x+\gamma} \frac{dx^{'}}{2\gamma}P(x^{'},t), & \mbox{if } \gamma\leq x \leq 1-\gamma,\\
\\
\delta(x-1) \int_{1-\gamma}^{1}
dx^{'}\frac{-1+\gamma+x^{'}}{2\gamma}P(x^{'},t)\\
+\int_{x-\gamma}^{1} \frac{dx^{'}}{2\gamma}P(x^{'},t), & \mbox{if } x \geq 1-\gamma.
\end{array} \right.
\label{G}
\end{equation}

For small values of $\gamma$ the boundary effects become less
important, and the second case in (\ref{G}) applies in the
majority of cases. In addition, for $\gamma$ small enough, this
term can be approximated as
\begin{equation}
\int_{x-\gamma}^{x+\gamma} \frac{dx^{'}}{2\gamma}P(x^{'},t)
 \approx
P(x,t)+\frac{\gamma^2}{6}\frac{\partial^2 P(x,t)}{\partial x^2}+
\ldots
\end{equation}
so that the diffusion term, i.e. the term proportional to $m$
in the right-hand side of Eq.~(\ref{eq:defuantdiff}) becomes of
the form $\frac{D}{2}\frac{\partial^2 P(x,t)}{\partial x^2}$,
where one finds again the diffusion coefficient
$D=m\gamma^2/3$.

\begin{figure}
\centering
\mbox{\epsfig{file=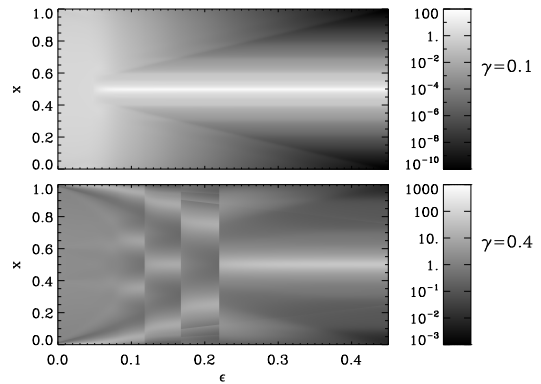,width=\linewidth,clip=,angle=0}}
\hfill
\caption{\small Plot of the asymptotic probability density $P(x,t\rightarrow\infty)$, coded
in logarithmic grey levels, as a
function of $\epsilon$, obtained after a numerical
integration of Eq.~(\ref{eq:defuantdiff}) starting with a flat
distribution. The top panel is for $\gamma=0.1$, and the bottom panel for
$\gamma=0.4$. In both cases we used $m=0.1$ and an integration time $t=5\times10^{4}$.}
\label{fig4}
\end{figure}

We have solved numerically the master equation (\ref{eq:defuantdiff}) for the
distribution $P(x,t)$ starting from an initial condition
representing an uniform distribution in opinion space, i.e.
$P(x,t=0)=1$ for $x\in[0,1]$ and $P(x,t=0)=0$ otherwise. For
$m=0$ it is well known that the distribution
$P_{\infty}(x)=\lim_{t\to\infty} P(x,t)$ is a sum of
delta-functions located at particular points
\cite{lorenz,redner}. However, this is not the case for the
full Eq.~(\ref{eq:defuantdiff}) with $m>0$: the final
distributions are no longer made up of delta-functions and their
final shape strongly depends on the particular values of
$\gamma$, $\epsilon$, and $m$. We have found that for small
values of $\gamma$ only one opinion cluster, centered around
$x=0.5$, remains for almost any value of the parameter
$\epsilon$ (see Fig.~\ref{fig4}a). We interpret this as a
consequence of the dynamics discussed in section \ref{sec:MCS},
where we showed that although several clusters were initially
formed, collisions reduced their number and a single one was
expected to survive at long times. We stress that the single
cluster in Fig.~\ref{fig4}a is only obtained at very long
times. At shorter times several clusters are present in the
solution of the master equation starting from a flat initial
condition (see for example Fig.~\ref{fig5}). The width of the
cluster in this master equation description should be related
to the dispersion seen in the Monte Carlo simulations (see Fig.
\ref{fig1}) at a given time, and not to the amplitude of the
diffusive cluster wandering, since the master equation
description is expected to be accurate as $N\rightarrow\infty$,
a limit in which cluster Brownian motion becomes frozen [see
(Eq.~\ref{CM}) and section \ref{subsec:MEMC}].

For large values of $\gamma$
the situation is rather different: as in the model with
free-will \cite{pineda}, for small $\epsilon$ several opinion
clusters form, which do not coalesce. Asymptotic states as a
function of $\epsilon$ are displayed in Fig.~\ref{fig4}b.
Different cluster bifurcations, as in the original Deffuant
model \cite{lorenz} and in the case of free-will \cite{pineda}
are seen.

An important feature is that, for all values of $\gamma$
considered, the clusters become less defined below a critical
value of $\epsilon$ (for large $\gamma$ one can always observe
however the two large clusters at $x=0,1$ arising from the
adsorbing boundary conditions). This point will be further
addressed in section \ref{sec:orderdisorder}.

\subsection{Master equation description versus Monte Carlo realizations}
\label{subsec:MEMC}

\begin{figure}
\centering
\mbox{\epsfig{file=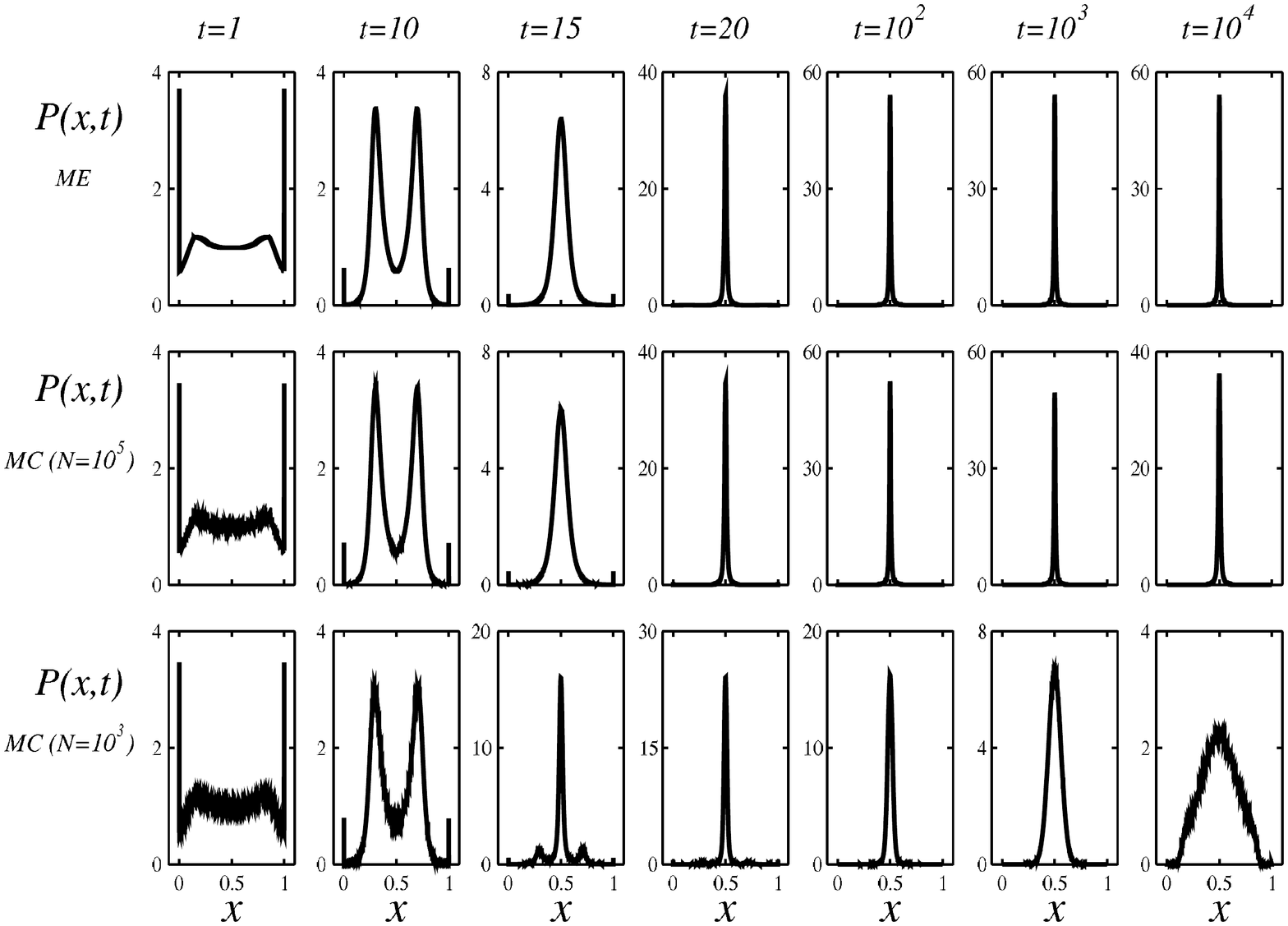,width=0.9\linewidth,clip=,angle=270}}
\mbox{\epsfig{file=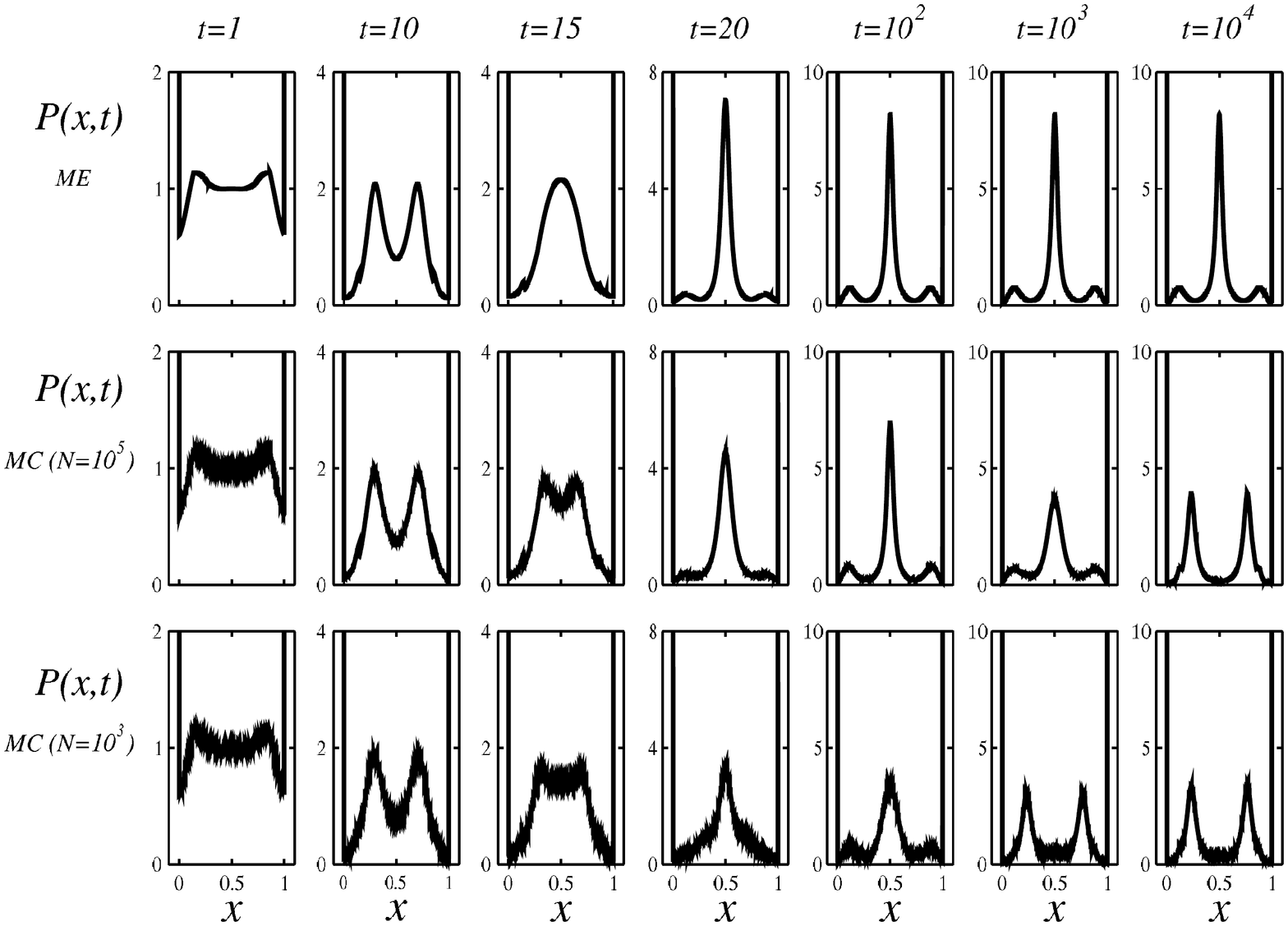,width=0.9\linewidth,clip=,angle=270}}
\hfill
\caption{\small Probability distribution function $P(x,t)$, for
intermediate time steps, from Monte-Carlo (MC) simulations
(histograms binned with bin size $\Delta x=5\times 10^{-4}$) for
two different system sizes $N=10^3$ and $N=10^5$, and the master
equation (ME) integrations of Eq.~(\ref{eq:defuantdiff}) starting
with a flat distribution with $\epsilon=0.28$ and $m=0.1$. The Monte Carlo
distributions are an average over $10^5$ realizations for $N=10^3$
and $10^4$ realizations for $N=10^5$. The top panels present
results with $\gamma=0.1$ and the bottom panels with
$\gamma=0.4$.}
\label{fig5}
\end{figure}

Monte Carlo simulations with a finite number $N$ of individuals
are not always faithfully represented by the solutions of the
master equation, which implicitly assumes that
$N\rightarrow\infty$, in addition to considering correlations
between individuals only in an approximate way. In
Fig.~\ref{fig5} we plot the time evolution of the probability
coming from Monte Carlo simulations (an average of the binned
opinion distribution over a large number of realizations is
performed, see caption) and the results from the master
equation in the case $m=0.1 $ and $\epsilon=0.28$, which is
close to a bifurcation from one to two big clusters in the
diffusionless model (see \cite{deffuant1} for details). The top
panels correspond to $\gamma=0.1$ and the bottom panels to
$\gamma=0.4$. It can be seen that, although the Monte Carlo
simulation and the master equation agree initially very well,
they start to deviate after a time that depends on the number
of individuals $N$: the larger $N$, the longer the time for
which the Monte Carlo simulations are faithfully described by
the master equation.

In the case $\gamma=0.1$, while the numerical solution of the
master equation and Monte Carlo simulations both reach at long
times a distribution with a single large maximum (large
cluster) at $x=0.5$, in the Monte Carlo simulations the width
of this steady distribution is larger for small particle number
$N$. One can understand this by noticing that the Monte Carlo
result is in fact the average over a large number of
realizations, and each one of them consists on a single cluster
whose location fluctuates widely (see Fig. \ref{fig1}). The
fluctuations in the location of the center of mass of the
cluster are reduced for increasing $N$ (see section
\ref{sec:MCS}), so that only the natural width of the cluster
will show up for increasing $N$, the range in which the master
equation description is expected to be accurate. For
$\gamma=0.4$, we recover the same type of behavior already
reported in our previous work concerning unbounded jumps in
opinion space \cite{pineda}. One can see in the bottom panel of figure \ref{fig5} that while the numerical solution of the master
equation tends at long times to a unimodal steady-state
distribution with a single large peak (surrounded by two small
ones), the Monte-Carlo simulations end up at long times in a
bimodal one with two peaks. At intermediate times both types of
solution agree, the larger the value of $N$ the longer the time
of agreement. A similar discrepancy between the Monte-Carlo and
the master equation results also appeared in our previous study
concerning unbounded jumps\cite{pineda}: starting from a
uniform initial condition, a unimodal steady state is reached
by the master equation but a bimodal distribution is the one
reached instead in Monte-Carlo simulations. The point is that
both distributions (the unimodal and the bimodal) are
stationary solutions of the master equation, but the unimodal
solution is metastable: a perturbation would take the system
out of this solution towards the bimodal distribution. The
perturbation needs to break the left-right symmetry (or
$x\to1-x$) of the problem which is present for a uniform
initial condition. In the case of the Monte-Carlo simulations
the perturbation is induced by the unavoidable finite-size
fluctuations, thus explaining the discrepancies. In the case of
the numerical integration of the master equation, this kind of
perturbation appears if one is not careful enough and
introduces, for example, round-off numerical errors that do not respect the
above-mentioned symmetry.

\section{Order-disorder transitions}
\label{sec:orderdisorder} As Fig.~\ref{fig4} shows, for
$\epsilon$ smaller than a critical value (which depends on $m$
and $\gamma$) the probability distribution becomes blurred such
that the maxima of the distributions are not evident, implying
the inhibition of cluster formation and the establishment of a
more homogeneous opinion distribution (although there is always
some inhomogeneity close to the boundaries, specially for large
$\gamma$). A similar effect can be observed with Monte-Carlo
simulations under adsorbing boundary conditions and can be
described in terms of an order-disorder transition: order
identified with the state with well defined opinion clusters
and disorder identified with the state without clusters.

To identify in a more quantitative way this order-disorder
transition, we use the so-called cluster coefficient $G_{M}$
\cite{pineda,vulpiani}. Its definition starts by first dividing
the opinion space $[0,1]$ in $M$ equal boxes and counting the
number of individuals $l_i$ which, at time step $n$, have their
opinion in the box $[(i-1)/M,i/M]$. The value of $M$ must not
be so large that particles are artificially considered to be
part of a single cluster, nor so small that statistical errors
are large within one box. We choose $M=100$. One next defines
an entropy $S_M=-\sum_{i=1}^{M}\frac{l_i}{N}\ln\frac{l_i}{N}$,
and the cluster coefficient \cite{vulpiani}
\begin{equation}
G_M=M^{-1}\left\langle e^{\overline S_M}\right\rangle,
\label{eq:CC}
\end{equation}
where the over-bar denotes a temporal average in steady
conditions and $\langle \cdot\rangle$ indicates an average over
different realizations of the dynamics. Note that $1/M\le
G_M\le 1$. Large values, $G_M \approx 1$, indicate that the
opinions are evenly distributed along the full opinion space (a
situation identified with disorder), while small values of
$G_M$ indicate that opinions peak around a finite set of major
opinion clusters (a situation identified with order).

\begin{figure}[htp]
\centering
\mbox{\epsfig{file=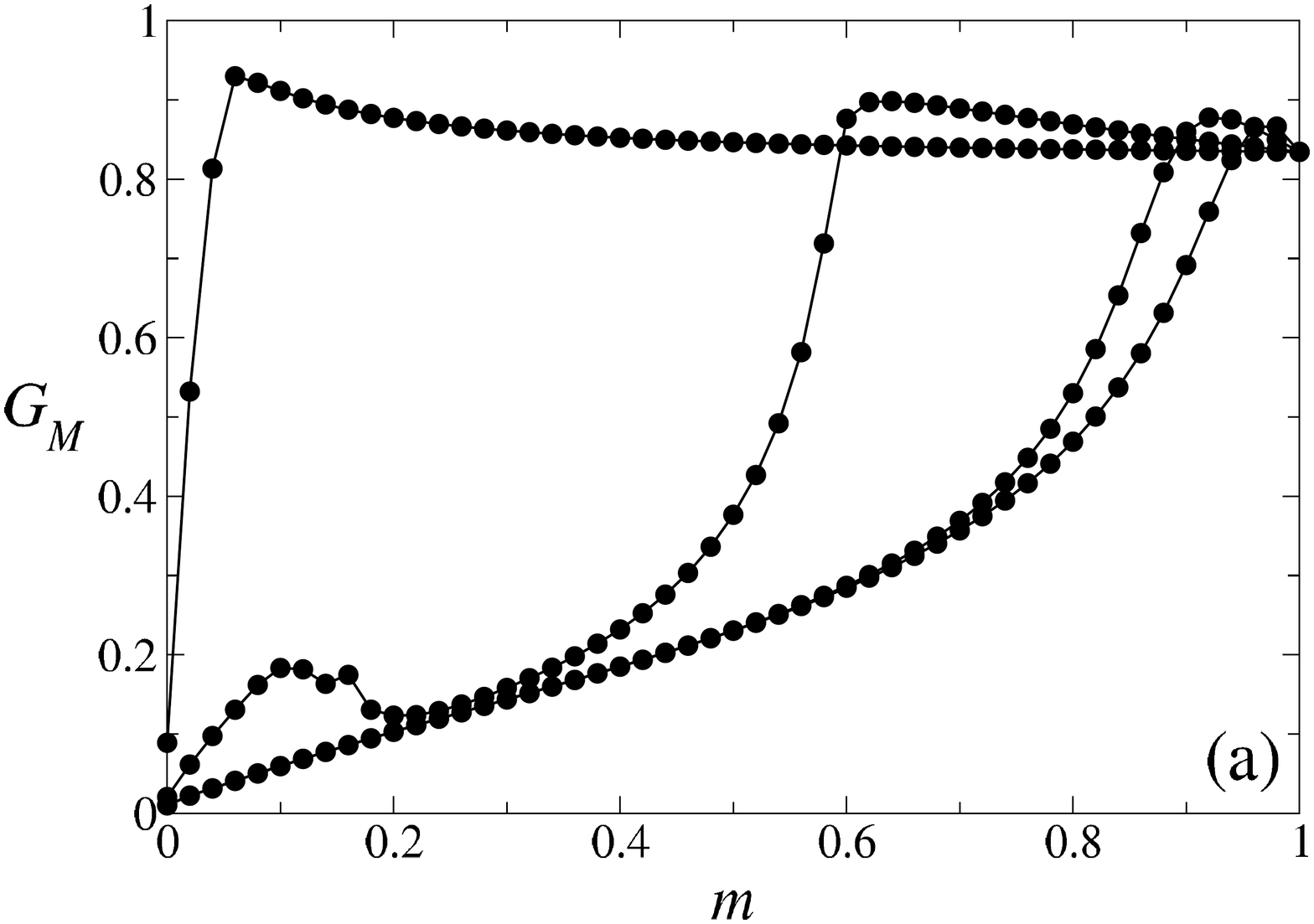,width=0.7\linewidth,clip=,angle=270}}
\mbox{\epsfig{file=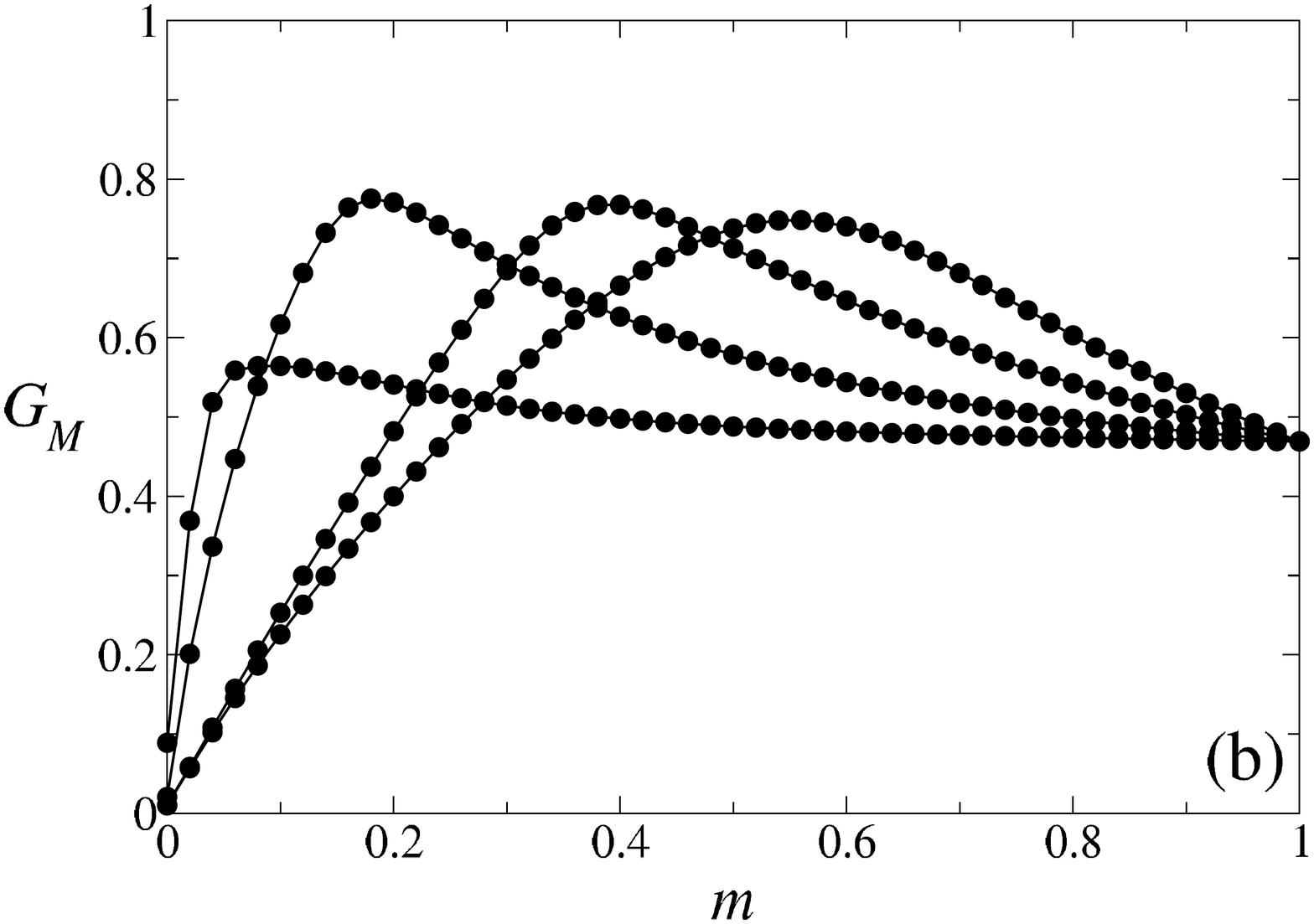,width=0.7\linewidth,clip=,angle=270}}
\hfill
\caption{\small Opinion cluster coefficient $G_{M}$ versus $m$ for
$\gamma=0.1$ (a) and $\gamma=0.4$ (b) obtained from Monte Carlo simulations
with $N=10^{4}$ and adsorbing boundary conditions. The values of the confidence parameter are
$\epsilon=0.05$, $0.2$, $0.4$, and $0.6$, from left to right
(dots; the solid line is a guide to the eye). We will define the
transition from order to disorder
as the location $m_c$ of the absolute maximum value of $G_M$.}\label{fig6}
\end{figure}

In Fig.~\ref{fig6}($a$) we plot $G_M$ as a function of $m$ for
$\gamma=0.1$ and different values of $\epsilon$ as obtained in
the Monte Carlo simulations using adsorbing boundary
conditions. For this value of $\gamma$, $G_M$ reaches an
absolute maximum value close to $1$ (corresponding to a
completely unstructured state) and then decreases
monotonically. The adsorption by the extreme opinion values $0$
and $1$ prevents the formation of a fully homogeneous state
(which will give $G_M=1$) as two opinion clusters are formed at
the extrema of the opinion space, whereas the rest of the
opinion space remains more or less homogeneously populated.
Fig.~\ref{fig6}($b$) shows a similar behavior in the case
$\gamma=0.4$, but in this case the number of individuals whose
opinion is adsorbed by the extremes is larger. Therefore, $G_M$
saturates farther away from its maximum possible value $1$. We
will define the transition from order to disorder as the
location $m_c$ of the absolute maximum value of $G_M$. In the
next subsection we will explain this transition via a simple
linear stability analysis that turns out to be very accurate,
in particular, for small values of $\gamma$.

\subsection{A linear stability analysis}
\label{sec:linear} Although the transition to cluster formation
is a nonlinear process, one can still derive approximate
analytical conditions for the existence of cluster formation as
a function of the control parameters by performing a linear
stability analysis of the unstructured solution of
Eq.~(\ref{eq:defuantdiff}). This is
greatly simplified if one neglects the influence of the
boundaries and assumes that there are periodic boundary
conditions at the ends of the interval $[0,1]$. This would be a
reasonable approximation for describing the distribution far
from the boundaries. We expect this approximation to be valid
for not too large $\gamma$ or $\epsilon$ since, as seen
for example in Fig.~\ref{fig4}, there is not much structure
near the edges of opinion space in those cases. Under periodic
boundary conditions, the homogeneous configuration $P_h(x)=1$
is the unstructured steady solution of the master equation
which results from Eqs.~(\ref{eq:defuantdiff}-\ref{G}) without
the contribution of the boundary terms:
\begin{eqnarray}
  \frac{\partial P(x,t)}{\partial t} & = & (1-m)[4\int_{|x-x_{2}|<\epsilon/2}dx_{2}P(2x-x_{2},t)P(x_{2},t) \nonumber \\
  & & - 2P(x,t)\int_{|x-x_{2}|<\epsilon}dx_{2}P(x_{2},t)] \nonumber  \\
  & & + m\left[\int_{x-\gamma}^{x+\gamma}\frac{dx'}{2\gamma}P(x't)-P(x,t)\right].
 \label{eq:deffuantdiffapprox}
\end{eqnarray}

To analyze the stability of the homogeneous solution $P(x)=1$
we write $P(x,t)=1+A_qe^{iqx+\lambda_q t}$, where $q$ is the
wave number of the perturbation, $\lambda_q$ its growth rate
and $A_q$ the amplitude. Introducing this ansatz in
Eq.~(\ref{eq:deffuantdiffapprox}) we find the dispersion
relation
\begin{eqnarray}
\lambda_q   & = & 4\epsilon(1-m)\left[\frac{4\sin(q\epsilon/2)}{q\epsilon}-\frac{\sin(q\epsilon)}{q\epsilon}-1\right] \nonumber \\
  & & + m\left[\frac{\sin(q\gamma)}{q\gamma}-1\right].
\label{eq:dispertionrelaA}
\end{eqnarray}

\begin{figure}
\centering
\mbox{\epsfig{file=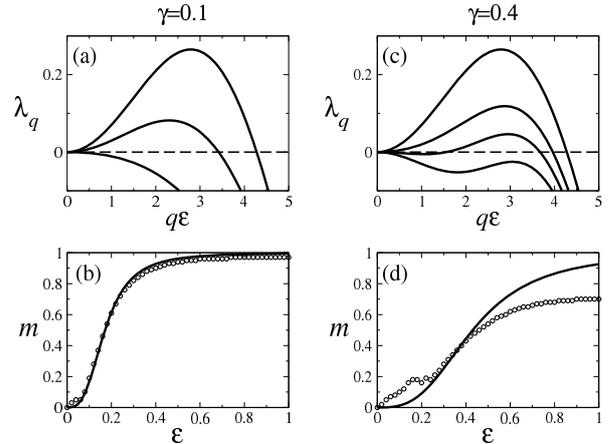,width=0.8\linewidth,clip=,angle=270}}
\hfill
\caption{\small (a) Growth rate, equation~(\ref{eq:dispertionrelaA}), of a perturbation to
the homogeneous state as a function of $q\epsilon$ for
$\epsilon=0.23$, $\gamma=0.1$, and $m=0.0$, $0.4$, and $0.8$, from
top to bottom. (b) Phase diagram on the plane ($\epsilon,m$) as
predicted by the linear stability analysis with $\gamma=0.1$
(solid line) is compared with Monte Carlo simulations using adsorbing boundary conditions
and $N=10^{4}$ (open dots). Disordered states are above the lines
and ordered (clustered) ones below. (c) Growth rate as a
function of $q\epsilon$ for $\epsilon=0.23$, $\gamma=0.4$, and
$m=0.0$, $0.1$, $0.15$ and $0.2$, from top to bottom. (d) Phase
diagram on the plane ($\epsilon,m$) as predicted by the linear
stability analysis with $\gamma=0.4$ (solid line) is compared with
Monte Carlo simulations using adsorbing boundary conditions and $N=10^{4}$
(open dots). Disordered states are above the lines
and ordered (clustered) ones below.}\label{fig7}
\end{figure}

We note that, for small $\gamma$ the second term, proportional
to $m$, becomes $-m(\gamma q)^2/6$ as corresponding to the
expected diffusive behavior. The top panels of Fig.~\ref{fig7}
show this dispersion relation as a function of $q\epsilon$ for
several values of $m$ with $\gamma=0.1$ and $0.4$. If there
exists a wavenumber $q$ for which $\lambda_q>0$, then the
unstructured uniform state is unstable and cluster formation
will be possible. It is not possible to find closed expressions
for the wavelength $q_{max}$ giving the maximum growth rate
$\lambda_{q_{max}}$, or the critical values $m_c$ and
$\epsilon_c$ defining the regions where the homogeneous state
is stable for fixed $\gamma$, but all of this can be readily
obtained numerically. Approximate analytical expressions can be
obtained expanding $ \lambda_q$ in powers of $q$:
\begin{eqnarray}
\lambda_q   & = & \left[-\frac{m\gamma^2}{6}+\frac{(1-m)}{3}\epsilon^{3}\right]q^2 \nonumber \\
  & & + \frac{1}{120}\left[m\gamma^4-\frac{7(1-m)\epsilon^5}{2}\right]q^4+O(q^6).
\label{eq:dispertionrelaB}
\end{eqnarray}

In the case in which the $q^4$ term remains negative (which
occurs if the $m\gamma^4$ term remains smaller than the one
containing $\epsilon$), the change of sign of the $q^2$ term
identifies $\epsilon_{c} =
\left[\frac{m\gamma^2}{2(1-m)}\right]^{1/3}$ as the value of
$\epsilon$ below which an unstructured configuration remains
stable. Alternatively, for fixed $\epsilon$ we find
$m_c=\frac{2\epsilon^{3}}{2\epsilon^{3}+\gamma^2}$, the
critical value above which clusters will disappear. Within this
approximation the expression for the fastest growing mode near
the order-disorder transition is:
\begin{equation}
q_{max} \approx
\sqrt{\frac{120}{7\epsilon_c^3-4\gamma^2\epsilon_c}}\left(\epsilon-\epsilon_{c}\right)^{1/2}.
\label{eq:fastesmode}
\end{equation}

We stress that all these expressions following Eq.
(\ref{eq:dispertionrelaB}) are only valid for the case in which
the appearance of positive values of $\lambda_q$ when varying a
parameter occurs first at values of $q$ close to zero,
corresponding to a long-wavelength instability. As seen in Fig.
\ref{fig7} this is not always the case for large values of
$\gamma$, for which the transitions have to be obtained
numerically. In fact, we note that the change in
behavior observed in the simulations from a tendency to
coarsening to the establishment of a robust periodic pattern,
occurring for increasing $\gamma$, seems to be correlated with
the change in the character of the linear instability, from
long-wavelength to finite wavelength, this last one being the
one occurring in the free-will model in \cite{pineda}. A more
detailed discussion of the relationship between the linear
dynamics and the nonlinear long-time states is however beyond
the scope of this paper. Comparison of the predicted
order-disorder transition with Monte Carlo simulations is
performed in the bottom panels of Fig. \ref{fig7}. We plot the
critical value $m_{c}(\epsilon)$ as obtained from the cluster
coefficient $G_{M}$ (as stated before, $m_c$ is defined as the
value at which $G_M$ is maximum) under adsorbing boundary
conditions together with the critical lines predicted by the
linear stability analysis. We see in the figure
that there is a very good agreement between theory and
simulations for $\gamma=0.1$, and a worse correspondence,
although still qualitatively correct, for $\gamma=0.4$. Thus we
can say that the arrest of cluster formation arises because the
random jumps stabilize the homogeneous opinion distribution.

\section{Summary and conclusions}
\label{sec:conclusions} We have studied the continuous opinion
model by Deffuant {\sl et al.} when one adds diffusion to the
dynamics. More precisely, we have modified the evolution rules
by including the probability that  individuals change their
opinion to another value randomly chosen inside an interval
centered around the current opinion. Our aim is to include the
effects of the unavoidable elements of randomness always
present in human decisions. When the typical size of the random
jumps is large, this model approaches a previous minimalistic
model introduced by us \cite{pineda}.

The final collective states depend on the system size $N$, the
confidence parameter $\epsilon$ as well as on the typical size
of the random jumps $\gamma$ and their probability $m$. While
our numerical results consider the more natural adsorbing
boundary conditions in which opinions beyond the limits of the
allowed interval  are set to the extreme values $0$ or $1$,
some analytical calculations are carried out in the case of
periodic boundary conditions, which are simpler from the
mathematical point of view.

The first observation one extracts from the Monte Carlo
simulations is the dispersion in the opinions within otherwise
well defined opinion clusters. We believe that this effect
comes closer to realistic situations in which a population
splits into different groups, but the opinions of individuals within each
group are not identical to each other. The detailed dynamics of
those clusters depends strongly on the average size of the
random jumps $\gamma$. For small values of $\gamma$, the center
of mass of each cluster performs a random walk through the
whole opinion space with an effective diffusion coefficient
that scales with $1/N$. As a consequence, opinion clusters
start to collide and merge in a coarsening process that leads
to a single large cluster at very long times. For large values
of $\gamma$, the mobility of clusters is reduced and several
opinion clusters can appear. Thus, a small diffusion mobility
favors opinion consensus.

We have derived a master equation for the probability density
function $P(x,t)$ which determines the individuals density or
distribution in the opinion space. Numerical integration of
this equation starting from uniform initial conditions reveals
that for small $\gamma$ only one opinion cluster centered
around $x=0.5$ is formed for almost any value of the parameter
$\epsilon$, as observed in the Monte Carlo simulations. For
large $\gamma$, a sequence of  bifurcations can be obtained in
the opinion space. This is similar to what was found in the
original Deffuant {\sl et al.} model or in our previous
modification \cite{pineda,redner}. We have also found that when
diffusion is included, the asymptotic steady-state probability
distributions reached by Monte-Carlo simulations are not always
well represented by the ones obtained from the master equation
dynamics starting from the same symmetric initial condition.
The Monte Carlo simulations and the master equation agree
initially, but start to deviate after a time that depends on
the number of individuals $N$: the smaller the size, the
earlier the deviation occurs. For example, it is possible to
observe in the simulations bistability between one state with
only one cluster (full consensus amongst the entire population)
and another state with  two clusters (lack of consensus or
polarization in the opinions), whereas this does not appear in
the master equation. We attribute this difference to the
inherent fluctuations arising from the finite number of
individuals of the Monte Carlo simulations. Of course, for
practical applications, the number of individuals will be
always finite and hence the predictions of the Monte Carlo
simulations should be more relevant than those of the master
equation.

An order-disorder transition to cluster formation induced by
diffusion has been characterized using the so-called cluster
coefficient $G_{M}$. Large $\gamma$ or $m$, or small
$\epsilon$, lead to disordered (quasihomogeneous or
non-clustered state). The value of $G_{M}$ in the disordered
state strongly depends on the parameter $\gamma$. For small
$\gamma$, $G_M$ reaches a maximum value close to $1$. The
presence of individual's opinions adsorbed at the extremes $0$
and $1$ does not disturb significantly the formation of an
otherwise homogeneous state. For $\gamma$ large enough, the
number of opinions adsorbed by the extremes increases and $G_M$
becomes more sensitive to $m$ and saturates far away from its
maximum value $1$.

We have presented a linear stability analysis that assumes
periodic boundary conditions at the ends of the $[0,1]$
interval. This analysis allows us to derive approximate
conditions for opinion cluster formation as a function of the
relevant parameters of the system. We have found a good
qualitative agreement between the linear stability analysis and
the numerical simulations using adsorbing boundary conditions
and small values of $\gamma$. For large values of $\gamma$ the
agreement is only qualitatively correct. Of course, the pattern
selection of this model is, with diffusion and without it,
intrinsically a nonlinear phenomenon and obtaining the exact
critical conditions for opinion group formation remains a
challenge.

Our work shows the impact of diffusion of opinions and
finite-size effects on the dynamics of continuous opinion
formation \cite{tt2007}. We want to emphasize that the
incorporation of random perturbations in opinion dynamics
induces novel and interesting phenomena and deserves to be
explored in more detail in future works.

\section*{Acknowledgments}
We acknowledge the financial support of project FIS2007-60327
from MICINN (Spain) and FEDER (EU) and project
FP6-2005-NEST-Path-043268 (EU). M. P. is supported by the
Belgian Federal Government (IAP project ''NOSY: Nonlinear
systems, stochastic processes and statistical mechanics'').

It is an honor to dedicate this paper to Pierre Coullet on the occasion of his 60th birthday.


\begin{thebibliography}{}
\bibitem{castellano} C. Castellano, S. Fortunato, and V. Loreto, Rev. Mod. Phys. {\bf 81}, 591 (2009).
\bibitem{galam} S. Galam, Eur. Phys. J. B {\bf 25}, 403 (2002).
\bibitem{sznajd} K. Sznajd-Weron and J. Sznajd, Int. J. Mod. Phys. C {\bf 11}, 1157 (2000).
\bibitem{schweitzer} F. Schweitzer and J. Holyst, Eur. Phys. J. B {\bf 15}, 723 (2000).
\bibitem{lorenz} J. Lorenz, Int. J. Mod. Phys. C {\bf 18}, 1819 (2007).
\bibitem{deffuant1} G. Deffuant, D. Neu, F. Amblard, and G. Weisbuch, Adv. Compl. Syst {\bf 3}, 87 (2000).
\bibitem{deffuant2} G. Weisbuch, G. Deffuant, F. Amblard, and J. P. Nadal, Complexity {\bf 7}, 855 (2002).
\bibitem{deffuant3} G. Weisbuch, G. Deffuant, and F. Amblard, Physica A {\bf 353}, 555 (2005).
\bibitem{krause1} R. Hegselmann and U. Krause, J. Artif. Soc. Soc. Simul {\bf 5}, 2 (2002).
\bibitem{markowich} B. D\"uring, P. Markowich, J. F. Pietschmann, and M. T. Wolfram, Proc. R. Soc. A {\bf 465}, 3687 (2009).
\bibitem{axelrod} R. Axelrod, J. Conflict Res. {\bf 41}, 203 (1997).
\bibitem{granovetter} M. Granovetter, American Journal of Sociology {\bf 83}, 1420, (1978).
\bibitem{pineda} M. Pineda, R. Toral, and E. Hern\'andez-Garc\'{\i}a, J. Stat. Mech. {\bf P08001}, (2009).
\bibitem{redner} E. Ben-Naim, P. L. Krapivsky, and S. Redner, Physica D {\bf 183}, 190 (2003).
\bibitem{vulpiani} A. Puglisi, V. Loreto, U. Marini Bettolo Marconi, and A. Vulpiani, Phys. Rev. E {\bf 59}, 5582 (1999).
\bibitem{tt2007} R. Toral and J. C. Tessone, Comm. Comp. Phys. {\bf 2}, 177 (2007).
\end{thebibliography}
\end{document}